\documentclass[prl,aps,assume,superscriptaddress,reprint]{revtex4-2}
\usepackage{graphicx}
\usepackage{dcolumn}
\usepackage{bm}
\usepackage{amsmath}
\usepackage{natbib}
\usepackage{color}
\usepackage{lineno,hyperref}
\usepackage[normalem]{ulem}

\newcommand{\Rmnum}[1]{\expandafter\@slowromancap\romannumeral #1@}
\makeatother
\begin{document}
\title{\bf Spin-Valve-Like Magnetoresistance and Anomalous Hall Effect in Magnetic Weyl Metal Mn$_2$PdSn}

\author{ Arnab Bhattacharya }
\affiliation{Condensed Matter Physics Division, Saha Institute of Nuclear Physics, A CI of Homi Bhabha National Institute, 1/AF, Bidhannagar, Kolkata 700064, India}
\author{ Mohammad Rezwan Habib }
\affiliation{School of Physical Sciences, Indian Association for the Cultivation of Science, 2A and 2B Raja S.C. Mullick Road, Jadavpur, Kolkata 700 032, India}
\author{Afsar Ahmed}
\affiliation{Condensed Matter Physics Division, Saha Institute of Nuclear Physics, A CI of Homi Bhabha National Institute, 1/AF, Bidhannagar, Kolkata 700064, India}
\author{ Biswarup Satpati}
\affiliation{Condensed Matter Physics Division, Saha Institute of Nuclear Physics, A CI of Homi Bhabha National Institute, 1/AF, Bidhannagar, Kolkata 700064, India}
\author{ Samik DuttaGupta}
\affiliation{Condensed Matter Physics Division, Saha Institute of Nuclear Physics, A CI of Homi Bhabha National Institute, 1/AF, Bidhannagar, Kolkata 700064, India}
\author{ Indra Dasgupta }
\email{sspid@iacs.res.in}
\affiliation{School of Physical Sciences, Indian Association for the Cultivation of Science, 2A and 2B Raja S.C. Mullick Road, Jadavpur, Kolkata 700 032, India}
\author{ I. Das}
\email{indranil.das@saha.ac.in}
\affiliation{Condensed Matter Physics Division, Saha Institute of Nuclear Physics, A CI of Homi Bhabha National Institute, 1/AF, Bidhannagar, Kolkata 700064, India}

\begin{abstract}

Realization of noncentrosymmetric magnetic Weyl metals is expected to exhibit anomalous transport properties stemming from the interplay of unusual bulk electronic topology and magnetism. Here, we present spin-valve-like magnetoresistance at room temperature in ferrimagnetic Weyl metal Mn$_2$PdSn that crystallizes in the inverse Heusler structure. Anomalous magnetoresistance display dominant asymmetric component attributed to domain wall electron scattering, indicative of spin-valve-like behavior. \textit{Ab initio} calculations confirm the topologically non-trivial nature of the band structure, with three pairs of Weyl nodes proximate to the Fermi level, providing deeper insights into the observed intrinsic Berry curvature mediated substantial anomalous Hall conductivity. Our results underscore the inverse Heusler compounds as promising platform to realize magnetic Weyl metals/semimetals and leverage emergent transport properties for electronic functionalities.
\end{abstract}

\maketitle

In the landscape of condensed matter physics, there has been a fervent pursuit of discovering novel quantum materials, aiming to validate topological principles and unravel their associated exotic properties\cite{p88,p89,p56,p31}. Among these, Weyl semimetals/metals (WSM) appear as a distinct class, emerging in crystals with broken inversion ($\mathcal{P}$)\cite{p55,p90} or time-reversal ($\mathcal{T}$) symmetry\cite{p42,p43}, hosting bulk emergent Weyl fermions\cite{p14,p63} and surface Fermi arcs connecting Weyl nodes of opposite chirality\cite{p22,p54}. Notably the magnetic Weyl semimetals (MWSM) with broken-$\mathcal{T}$, offer an expanded realm to explore the interplay between topological ordering, magnetism and electron correlation compared to their non-magnetic counterparts\cite{p14,p43,p63}. In $\mathcal{P}$-conserved MWSMs, the broken-$\mathcal{T}$ induced persistent Berry curvature along with intrinsic magnetism results in remarkably large anomalous Hall conductivity (AHC) \cite{p28,p32,p54,p58,p4} and anomalous Hall angle\cite{p14,p43}, positioning MWSMs favourably for potential applications in spintronics. However, research in MWSMs has primarily focused on ferromagnets, where topological features are protected by mirror symmetry under conserved $\mathcal{P}$, subject to breakdown upon the inclusion of spin-orbit coupling (SOC). This limits the concurrent realization of real-space non-collinear magnetic ordering, like skyrmions\cite{p77,p53} or spin-valve effect\cite{p15,p16}, and non-trivial \textit{k}-space topology, hindering exploration of the novel magnetic responses and valuable insights this synergy may offer.

In this context, 4\textit{d}-transition-element-based 'inverse' Heusler alloys (iHA) present an intriguing platform with noncentrosymmetric crystal structures and multiple inequivalent magnetic sublattices\cite{p10}. The advantages of 4\textit{d}-transition metal-based alloys over extensively studied 3\textit{d} counterparts lie in reduced antisite disorder, resulting from differences in size and electronegativity with other constituent elements, and enhanced SOC effects. The inherent magnetism of iHA lifts the band spin degeneracy, while moderate $\mathcal{P}$-breaking preserves the topological nature of the Weyl nodes, shifting them to different energies from the Fermi level. This shift eliminates the possibility of \textit{nodal semimetals} while maintaining Weyl semimetal properties, like chiral anomaly and large AHC\cite{p62}. In real-space scenario, the multiple magnetic sublattices introduce an additional layer of intricacy. The ferrimagnetism in majority of cubic Mn-based iHA arises from the antiferromagnetic coupling between Mn atoms in the tetragonal environment and highly localized octahedrally coordinated Mn atoms\cite{p35,p36,p37}, exhibiting emergent phenomena like exchange bias \cite{p34} and the rare spin-valve effect\cite{p15,p16}, serving a novel backdrop to explore the relation of Weyl fermions and magnetism.

Here we present Mn$_2$PdSn as candidate Weyl metal demonstrating room temperature spin-valve-like magnetoresistance and substantial AHC, through combined experimental and theoretical study. Isothermal field-dependent longitudinal resistivity exhibits anomalous magnetoresistance (\textit{MR}) dominated by asymmetric component indicating rare spin-valve behaviour in stoichiometric systems. \textit{Ab initio} calculations confirm the presence of Weyl points close to the Fermi level offering deeper insights into the Berry curvature mediated intrinsic AHC. We present a nontrivial AHC in the compensated ferrimagnetic Mn$_2$PdSn by determining the magnetic space group which contains magnetic mirror planes. These results underscores iHA as a platform for simultaneous realization of non-trivial magnetic functionalities and electronic topology.

Polycrystalline ingots of Mn$_2$PdSn were prepared by a conventional arc-melting technique (see supplementary information (SI)\cite{R30}). The compound crystallized in iHA structure (space group $F\Bar{4}3m$) with a stoichiometric representation $XYX'Z$, where $X$, $Y$, $X'$ and $Z$ occupies the 4$c$, 4$d$, 4$b$ and 4$a$ Wyckoff sites respectively, across four interpenetrating FCC sublattices. Selected area electron diffraction pattern (SAED) along the [110] zone axis and inverse fast Fourier transformation of the high-resolution TEM image of the samples, revealing the high crystal quality (see Fig.SF1 of SI \cite{R30}).

\begin{figure}[t]
  \begin{center}
  \includegraphics[width=0.48\textwidth]{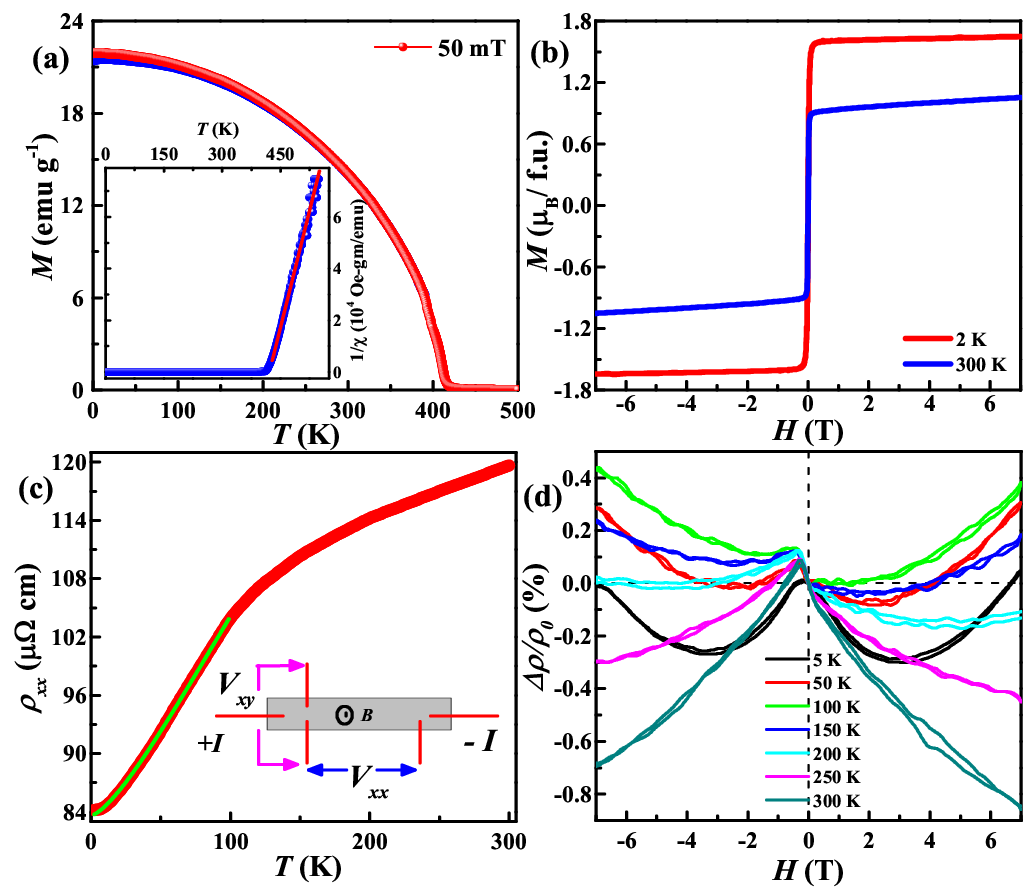}
  \end{center}
  \caption{(a) Temperature variation of magnetization under the applied field of 50 mT. The inset shows the inverse susceptibility curve versus temperature. (b) Isothermal magnetization $M(H)$ at 2 K and 300 K. (c) Temperature variation of $\rho_{xx}$ with the green lines showing the fitted data. Schematics show the sample configuration used for magnetotransport measurements. (d) Magnetic field dependance of isothermal magnetoresistance ($\Delta\rho/\rho_0$) at different temperatures. }
   \label{F1}
\end{figure}

\begin{figure}[t]
\begin{center}
\includegraphics[width=0.48\textwidth]{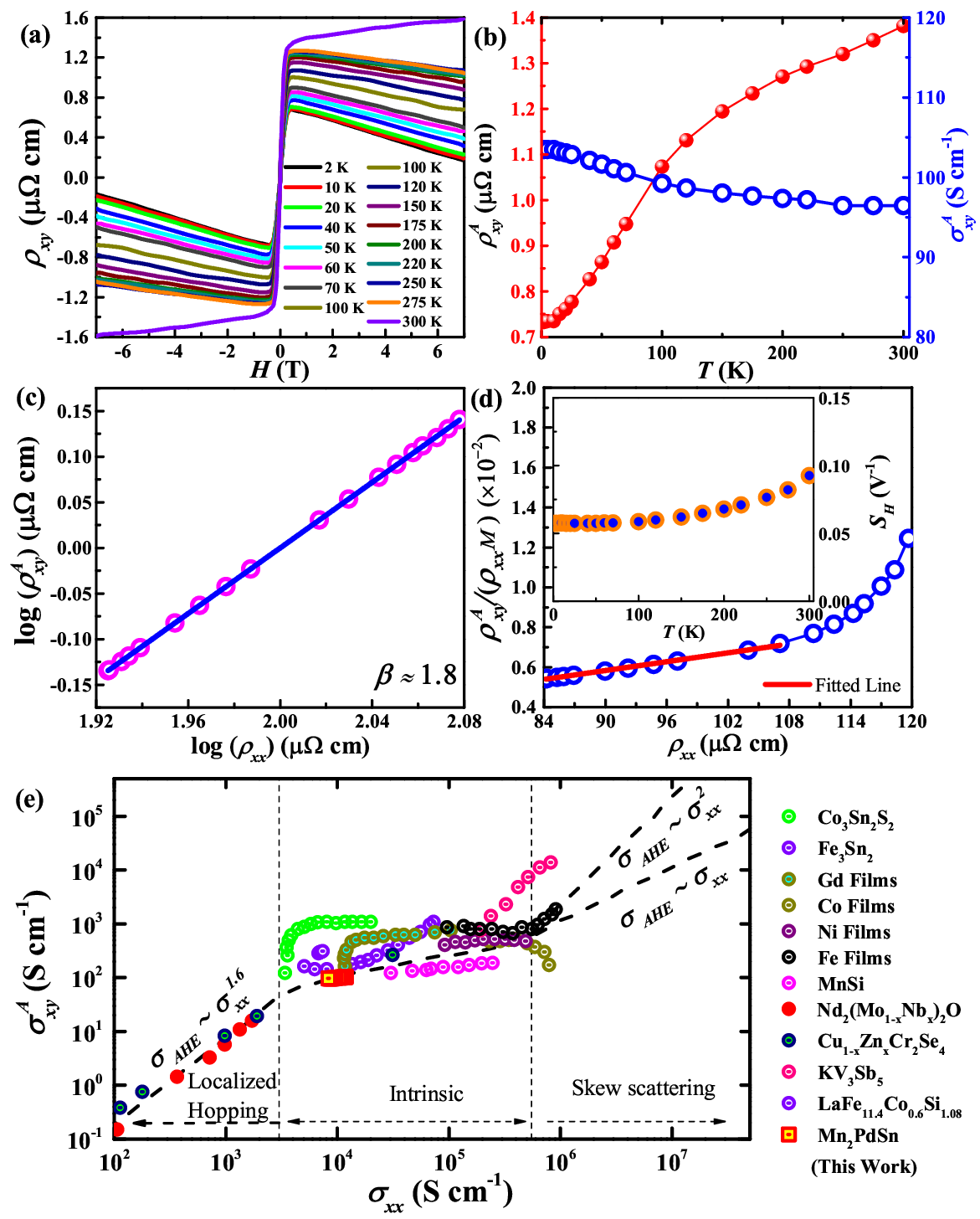}
\end{center}
\caption{(a) Isothermal transverse resistivity ($\rho_{xy}$) from 2-300 K (b) Temperature variation of Anomalous Hall resistivity, $\rho_{xy}^A$ and conductivity, $\sigma_{xy}^A$. (c) Plot of log($\rho_{xy}^A$) vs. log($\rho_{xx}$) with the slope $\alpha \approx 1.8$ (d) \textit{TYJ} scaling of Hall resistivity. Inset shows the \textit{T} variation of $S_{H} =\sigma_{xy}^A/M_S$. (e) Universal plot for $\sigma_{xy}^A$ with $\sigma_{xx}$ showing the intrinsic regime for Mn$_2$PdSn with earlier reported materials\cite{p40}.  }
\label{F2}
\end{figure}

Like the majority of Mn$_2$-based iHA, thermomagnetic measurements $M(T)$ under an applied field of 50 mT reveal a para to ferro(FM)/ferrimagnetic(FiM) transition above room temperature for Mn$_2$PdSn at $T_C=$ 410 K (Fig.\ref{F1}(a)). The susceptibility $\chi$ follows the Curie-Weiss (CW) law with $\theta_{CW}\approx$  421.5 K (inset of Fig.\ref{F1}(a)). The isothermal magnetization $M(H)$ at 2 K and 300 K reflects a soft FM/FiM nature with negligible hysteresis (Fig.\ref{F1}(b)). For iHA, the net magnetic moment follows the Slater-Pauling (SP) rule as $M_{sat} = (N_v - 24) \mu_B/$ f.u., where $N_v$ is the total number of valence electrons\cite{p10}.  Interestingly, from $M(H)$ isotherm at \textit{T} = 2 K, we obtain a spontaneous moment of 1.62 $\mu_B$/f.u., significantly lower than the SP rule estimate for complete FM ordering ($M_{sat}$ = 4 $\mu_B$/f.u, $N_v=$ 28). This difference is attributed to the inter-sublattice antiferromagnetic correlation between octahedrally coordinated Mn$_{4b}$ atoms in the Mn-Sn plane and Mn$_{4c}$ at the tetrahedral site in Mn-Pd sublattice, resulting in net FiM ordering\cite{p45,p35,p34,p36,p37}.

Figure.\ref{F1}.(c) shows the temperature variation of longitudinal resistivity $\rho_{xx}$ exhibiting a metallic behaviour. To elucidate the dominant scattering mechanism in $\rho_{xx}$, we fit $\rho_{xx}(T)$ between 2 and 100 K using the \textit{e-e} scattering $T^2$ dependence. However, a more accurate low-temperature fit incorporates the spin-fluctuating $T^{3/2}$ term, yielding $\rho_{xx} = \rho_{0} + aT^{3/2} + bT^{2}$, where $\rho_{0}$ is residual resistivity, and $a$ and $b$ are constants \cite{p18,p19}. The fitting results in $\rho_{0} = 83.615$ $\mu\Omega$cm, $a = 3.442 \times 10^{-2}$ $\mu\Omega$cmK$^{-1}$, and $b = 1.39 \times 10^{-3} \mu\Omega$cmK$^{-2}$, highlighting the prominent role of spin-fluctuation at low temperatures. This underscores that the interaction between conduction electrons and localized moments is not a weak perturbation but a strong electron-magnon interaction\cite{p61}.

The magnetic field dependence of \textit{MR} isotherms [$\Delta\rho/\rho_0 = \frac{\rho_{xx}(H)-\rho_{xx}(H=0)}{\rho_{xx}(H=0)}\times100\%$] typically exhibits symmetry concerning the applied $H$ direction. Conversely in Fig.\ref{F1}(d) Mn$_2$PdSn exhibits distinct asymmetry around $H = 0$ upon field reversal across the entire $T$ range, with the asymmetry along the initially applied field directon while the curve retraces when field is swept in opposite direction, reminiscent of magnetic-spin-valve behavior\cite{p15,p16}. The assymetric component of the \textit{MR} defined as, \textit{MR}$^{asm}$ = [\textit{MR}$(H)$ $-$ \textit{MR}$(-H)$]/2, showcase a persistent presence from 5 to 300 K (see Fig.SF3(b) of supplementary\cite{R30}). \textit{MR}$^{asm}$ closely mirrors the $M(H)$ curves, underscoring the predominant influence of the change in magnetic ordering at low $H$ within the magnetic domains on \textit{MR}$^{asm}$. Intuitively, \textit{MR}$^{asm}$ can be attributed to the domain wall induced spin-dependant scattering of conduction electrons. The electrons follow a least-resistive path stemming from unidirectional spins within FM clusters embedded in FiM matrix with oppositely orientated moments at the domain walls, experiencing frustrated state, acts as pinning centres. This magnetic arrangement results in two distinct resistivity states, reflected as the spin-valve behavior in MR upon $H$ reversal in low field region, sustained even at room temperature. To elucidate the temperature evolution of \textit{MR} (Fig.\ref{F1}(d)), we separate the symmetric component \textit{MR}$^{sym}$ = [\textit{MR}$(H)$ $+$ \textit{MR}$(-H)$]/2 (see Fig.SF3(a) of supplementary\cite{R30}). At lower $T$, \textit{MR}$^{sym}$ exhibits a positive slope for $H \ge$ 3T, attributed to the Lorenz force acting on the conduction electrons. In contrast, the negative \textit{MR}$^{sym}$ at high-$T$ originates from the suppression of spin scattering with the application of $H$\cite{p74}. Thus establishing a close correlation between magnetism and transport properties.

Now we turn to Hall transport data. Figure.\ref{F2}(a) shows the transverse resistivity $\rho_{xy}$ vs. $H$ at various $T$. For FM/FiM conductors, the $\rho_{xy}$ can be emperically expressed as, $\rho_{xy} = \rho_{xy}^O + \rho_{xy}^A$, where $\rho_{xy}^O = R_0H$ and $\rho_{xy}^A= R_SM$ are ordinary and anomalous Hall resistivity, respectively\cite{p28,p32,p54}. The negative slope establishes electrons as the majority charge carriers with carrier density $n_0 = 7.8 \times 10^{21}$ cm$^{-3}$ at $T$ = 2 K. $\rho_{xy}^A$ is obtained from high $H$ extrapolation to zero field, displaying an increasing trend with temperature (Fig.\ref{F2}(b)), contrary to the behavior of $M_s$. In a generalized picture, the AHE stems from either intrinsic mechanism mediated by reciprocal space Berry curvature or extrinsic side-jump(\textit{sj})/skew(\textit{sk}) scattering mechanisms\cite{p8}. To scrutinize the origin, we derive the AHC $\sigma_{xy}^A$, akin to obtaining $\rho_{xy}^A$, from total Hall conductivity $\sigma_{xy} \approx \rho_{xy}/\rho_{xx}^2$. $\sigma_{xy}^A$ exhibits nearly constant variation with $T$, with a value of $\sigma_{xy}^A \approx 103.5$ Scm$^{-1}$ at $T =$ 2 K, suggesting an intrinsic origin of AHE (Fig.\ref{F2}(b))\cite{p42,p43,p44}. In Fig.\ref{F2}(c), the linear relation between log($\rho_{xy}^A$) and log($\rho_{xx}$) is fitted using the relation $\rho_{xy}^A \propto \rho_{xx}^\alpha$ which yields $\alpha \approx 1.79(4)$, suggestive of the dominant contribution of intrinsic Berry curvature or extrinsic \textit{sj} mechanism in AHE \cite{p38}. To quantify these contributions, we adopt the {\it TYJ} scaling relation\cite{p39}, $\rho_{xy}^A/M\rho_{xx} = a + b\rho_{xx}$, where the first term $a$ stems from extrinsic \textit{sk} mechanism and the second term, $b = \rho_{xy}^A/\rho_{xx}^2$, relates directly with intrinsic AHC, $\sigma_{xy,int}^A$. From the scaling, a linear relation is expected between $\rho_{xy}^A/(M\rho_{xx})$ and $\rho_{xx}$, which prevails for $\rho_{xx} \leq 108$ $\mu\Omega$cm (corresponding to $T$ = 140 K) resulting in $a \approx - 8.1\times 10^{-4}$ and $b \approx 73.8$ S cm$^{-1}$ (Fig.\ref{F2}(d)). The $\sigma_{xy,int}^A$ accounts to $70\%$ of $\sigma_{xy}^A$ at $T$ = 2 K. At low temperatures with reduced phonon scattering, the $\sigma_{sj}^A$ entangles with $\sigma_{xy,int}^A$, posing a challenge for individual quantification due to the absence of a scaling framework. However, an estimation of the order of magnitude of $\sigma_{sj}^A$ \cite{p40} reveals it to be notably smaller than $\sigma_{xy,int}^A$, thereby confirming the Karplus-L$\ddot{u}$ttinger origin of AHC\cite{p8}.

Fig.\ref{F2}(e) shows the $\sigma^A_{xy}$ versus $\sigma_{xx}$ plot along with other known FM/FiM where it lies well within the limit for the intrinsic origin of AHE. To gauge the robustness of AHE, we employ the anomalous Hall factor $S_H$ (= $\sigma^A_{xy}/M_{S}$), quantifing the sensitivity of the magnitude of anomalous Hall current in regard to the magnetization. Inset of Fig.\ref{F2}(d) illustrates the variation of $S_H$ with temperature, and it remains fairly constant with $T$ at $S_H \approx 0.065$ V$^{-1}$. This signifies that the observed AHE is immune towards impurity scattering, prompting further investigation into the origin of the Berry curvature in \textit{k}-space through \textit{ab-initio} calculations.

\begin{figure}[t]
	\begin{center}
		\includegraphics[width=0.48\textwidth]{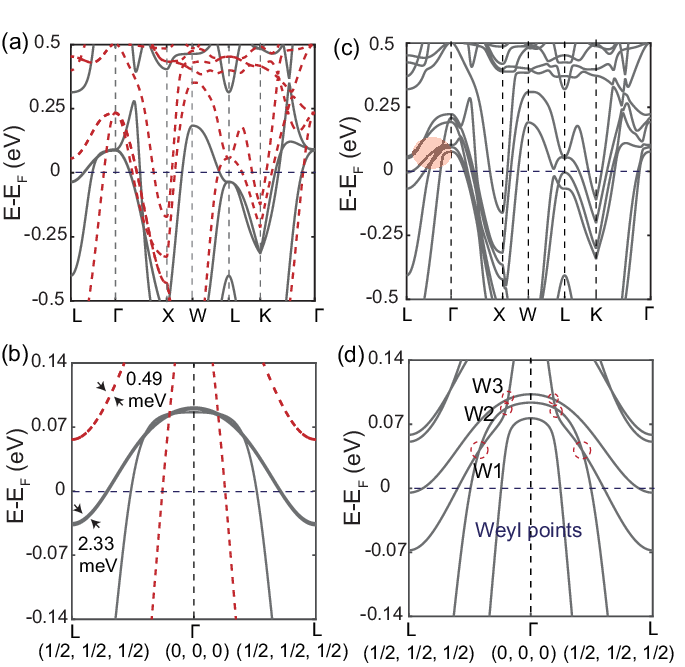}
	\end{center}
	\caption{(a) Spin-polarized band structure of Mn$_2$PdSn along high symmetry $k$ path for ferrimagnetic (FiM) configuration where up and down spin states are representated with full and dotted lines respectively.(b) The same band structure along \emph{L-$\Gamma$-L} path showing non-degenerate nature of the band and their crossings near the Fermi level. (c) Band structure including SOC. The highlighted area in (c) shows several band crossing points. (d) The same band structure along the \emph{L-$\Gamma$-L} path shows three pairs of Weyl points in the range of $\pm$100 meV above the Fermi level. The Fermi Level is set at 0 eV.}
	\label{F3}
\end{figure}

To understand the impact of magnetism on the electronic structure we have performed spin-polarized calculation in the frame work of Density functional theory (DFT) using Generalized gradient approximation (GGA) by considering the FM and FiM configuration of Mn spins. Our calculations reveal FiM configuration is energetically favourable in comparison to FM configuration. The net magnetic moment of Mn$_2$PdSn in the ferrimagnetic state is claculated to be 0.48$\mu_B$. The nearly equal and opposite magnetic moment of 4.54 $\mu_B$ and -4.44 $\mu_B$ hosted respectively by Mn$_{4b}$ and Mn$_{4c}$ atoms in the primitive unit cell of Mn$_2$PdSn results in a small net moment. The presence of net magnetic moment, although small, breaks the $\mathcal{T}$ symmetry. The spin resolved total and partial density of states for the FiM configuration depict the presence of finite states near Fermi level for both the spin channels which makes it metallic (Fig.SF5 of supplementary\cite{R30}).

The band structure for the FiM configuration along the high symmetry $k$-path for Mn$_2$PdSn is shown in Fig. \ref{F3}(a). Due to the tetrahedral crystal field, the valance band maxima at the $\Gamma$-point for both the majority and minority spin channels are threefold degenerate which are mainly composed of Mn-\emph{$t_{2}$} states. Along the \emph{$\Gamma$-L} direction, near the Fermi level, the three fold degenerate bands are expected to split into a doubly degenerate and a non-degenerate band, according to the irreducible representation (IRS) of $C_{3v}$ little group. However, due to the absence of $\mathcal{T}$ symmetry in the FiM ground state, the degeneracy is lifted (Fig.\ref{F3}(b)). Along \emph{$\Gamma$-L} several band crossings between these non-degenerate bands are observed above the Fermi level.

\begin{figure}[t]
	\centering
	\includegraphics[width=.49\textwidth]{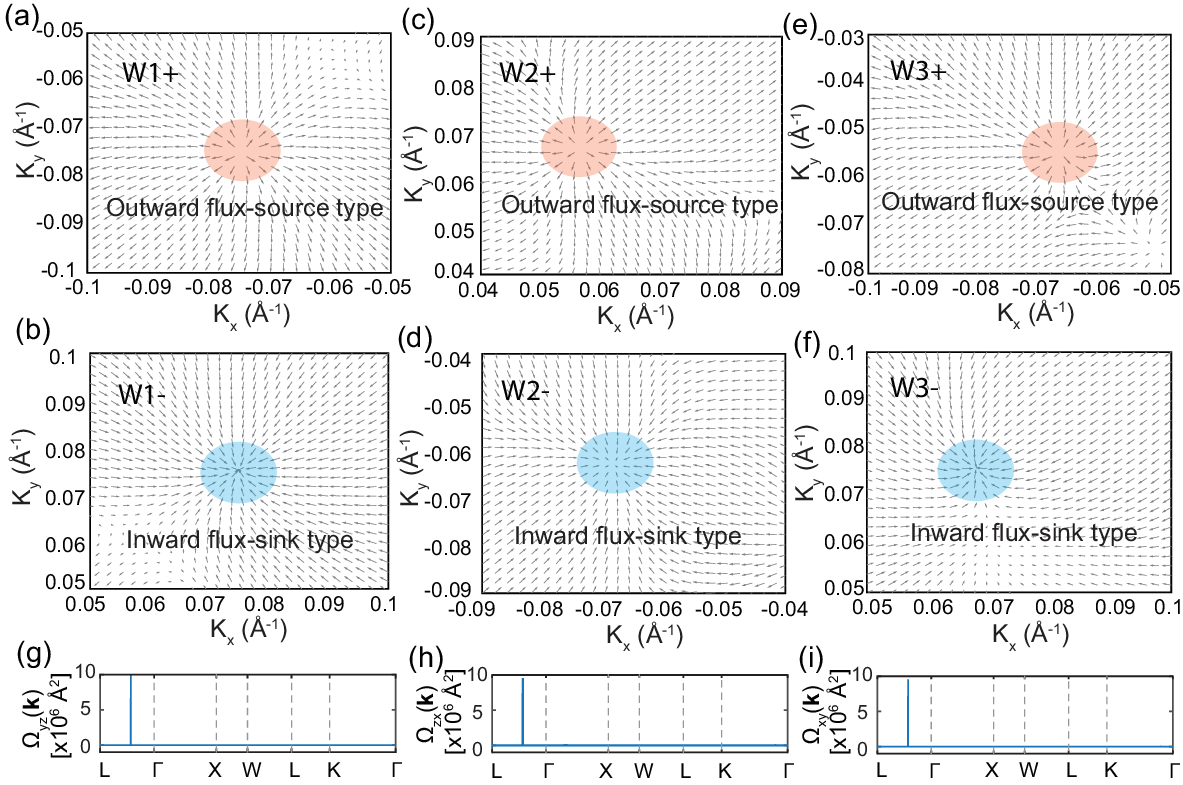}
	\caption{(a-f) Normalized Berry curvature distribution around the Weyl points with positive and negative chirality. In (a, c, e) source type assigned by $+$ symbol, which is indicated by outward arrows as highlighted by an orange ellipse. In (b, d, f) sink type assigned by the $–$ symbol is depicted by inward arrows as emphasized by blue ellipse. The Berry curvatures along the high symmetry $k$-paths are shown in (g-i).}
	\label{F4}
\end{figure}

Next, we have included SOC in our calculation. The calculated energy of Mn$_2$PdSn in FiM configuration including SOC where the magnetization is along [111] direction is found to be the lowest. Our computational results predict that Mn$_2$PdSn stabilizes in a noncollinear ferrimagnetic phase where the magnetization axis is along [111] direction. The band structure in the FiM configuration including SOC along [111] direction is shown in Fig. \ref{F3}(c), where in the presence of SOC the three fold degeneracy at the \emph{$\Gamma$} point is lifted, and several band crossings are observed along the path \emph{$\Gamma$-L}. The band structure along the path \emph{$\Gamma$-L} in a narrow energy range (see Fig.\ref{F3}(d)) identifies several doubly degenerate points (Weyl points) above the Fermi level. In the spin-orbit coupled band structure along \emph{L-$\Gamma$-L} path, three pairs of Weyl points are located about 100 meV above the Fermi level, and are distinctly seen in Fig.\ref{F3}(d). The precise position, Chern number and the chemical potential of these Weyl points are determined by the iterative Greens function approach as implemented in the Wannier Tools package \cite{p91} (Table. ST2 of SI \cite{R30}). To understand the texture of these Weyl points, we have calculated the normalized Berry curvatures which indicate the flux at the Weyl point.  The normalized Berry curvature shows that a Weyl point either acts as a source, where the flux comes outward from the point (e.g., W1+, W2+, W3+) or as a sink, where the flux is along the inward direction (e.g., W1-, W2-, W3-), of the Berry curvature, clearly showing the monopole characteristic with definite chirality as illustrated in Fig.\ref{F4}(a)-(f)

The Weyl points in the FiM structure are expected to generate finite Berry curvature resulting in transverse anomalous velocity in the electronic motion subsequently producing a large AHC thereby modifing the transport behaviour of the compound. The intrinsic AHC ($\sigma_{\alpha\beta}$) is expressed as the integral of the total Berry Curvature ($\Omega_{\alpha\beta}$) over the Brillouin zone of the crystal as\cite{p50,p51}, $\sigma_{\alpha\beta} = \frac{e^2}{\hbar}\int_{BZ} \frac{d^3\Vec{k}}{(2\pi)^3}\Omega_{\alpha\beta}(\Vec{k})$, where $\Omega_{\alpha\beta}(\Vec{k}) = \sum_{n}f_n(\Vec{k})\Omega^\gamma_{n}(\Vec{k})$ is the sum of the Berry curvatures $\Omega_{n,\alpha\beta}(\Vec{k})$ corresponding to the individual bands n, $f_n(\Vec{k})$ is the Fermi distribution function and indices ($\alpha$, $\beta$) denote the global cartesian coordinates\cite{p50,p52}. The Berry curvature can be cast into the form of a Kubo-like formula which is implemented in the WANNIER90 package \cite {p92} as, $\Omega_{n,\alpha\beta}(\Vec{k}) = -2i\hbar^2\sum_{m\neq n} \frac{\langle \Psi_{n, \Vec{k}}|v_\alpha|\Psi_{m, \Vec{k}}\rangle\langle \Psi_{m, \Vec{k}}|v_\beta|\Psi_{n, \Vec{k}}\rangle}{[E_m(\Vec{k}) - E_n(\Vec{k})]^2}$, where $\Psi_{n, \Vec{k}}$ is the Bloch function and $\Vec{v}$ is the velocity operator.

The magnetization direction along [111] alters the governing symmetry from cubic to trigonal \cite{p82,p83,p84}. The symmetry is described by the magnetic space group $R3m'$ (160.67), where the prime denotes the time-reversal symmetry. The resulting constraints imposed on the Berry curvature $\Omega_{\alpha\beta} = (\Omega_{yz}, \Omega_{zx}, \Omega_{xy})$ and therefore, on the AHC $\sigma_{\alpha\beta} = (\sigma_{yz}, \sigma_{zx}, \sigma_{xy})$ by the magnetic space group $R3m'$ can be determined. Notably, the magnetic point group $3m'$ is sufficient to determine the constraints \cite{p85,p86}. Here, it is worth mentioning that the magnetic mirror symmetry reverses the spin components perpendicular to the magnetic mirror plane and preserves the spin components parallel to the magnetic mirror plane \cite {p87}. Therefore, the application of the symmetry operation of the magnetic mirror planes $m'_{10\bar{1}}$, $m'_{01\bar{1}}$ and $m'_{1\bar{1}0}$ preserves the non-collinear ferrimagnetic configuration of Mn$_2$PdSn.

The above three magnetic mirror planes make all three components of the Berry curvature to be non-trivial hence, the AHC takes the form, $\sigma_{\alpha\beta} = (\sigma_{yz}, \sigma_{zx}, \sigma_{xy})$. Figure.\ref{F4}(g)-(i) shows the calculated Berry curvature along high symmetry $k$ path. The non-vanishing nature of the three components of the Berry curvature is in agreement with the magnetic symmetry of Mn$_2$PdSn compound. The peaks in the plot of Berry curvature along \emph{$\Gamma$-L} direction originate due to the negligible band separation between the band crossing point.

We have calculated three components of the AHC, $\sigma_{yz}$, $\sigma_{zx}$ and $\sigma_{xy}$ where x, y and z axis are considered along [100], [010] and [001] direction of the crystallographic axis. For the magnetization along [111] direction, the AHC tensor can be calculated using another basis where x, y, and z axis are considered along [$\bar{1}$10], [$\bar{1}$$\bar{1}$2] and [111] direction, respectively. The calculated AHC values (shown in Table.ST2 of supplementary\cite{R30}), are -58.2, -68.0, and -60.0 S cm$^{-1}$  for $\sigma_{yz}$, $\sigma_{zx}$ and $\sigma_{xy}$, respectively. In the transformed basis the dominant contribution is the $\sigma'_{xy}$ = -107.4 S cm$^{-1}$ is in good agreement with the experimentally obtained intrinsic contribution of the AHC value.

In conclusion, this comprehensive experimental and {\it ab-initio} investigation highlights the broken inversion symmetry in the iHA system as a conducive environment for realizing bulk Weyl singularities and exploring their associated emergent transport properties. Detailed magnetotransport measurements unveil asymmetric magnetoresistive behavior, primarily originating from domain wall scattering at applied low magnetic fields, resulting in a persistent spin-valve-like effect up to room temperature. Moreover, \textit{MR}$^{asm}$ exhibits similar temperature and magnetic field variations as magnetization, emphasizing the crucial role of magnetic ordering in the spin-valve effect. A quadratic scaling relation between $\rho^A_{xy}$ and $\rho_{xx}$ is observed, and the calculated AHC resulting from the Berry curvature produced by the three pair of Weyl nodes adjacent to the Fermi level closely agrees with the experimentally observed $\sigma^A_{xy,int}$, confirming the \textit{k}-space topology as the origin of AHE. Our results underscore the Mn-based iHA with late transition elements at the \textit{Y} position as a fertile playground for exploring the intricate interplay of non-collinear magnetism and non-trivial electronic band topology. This study further merits the possibility of exploring exotic magnetic functionalities of MWSMs with broken inversion symmetry, like topological spin-valve technology, owing to the strong correlation between magnetization dynamics and electromagnetic fields mediated by Weyl fermions, through experimental and theoretical investigation for potential applications.

\emph{\textbf{Acknowledgements}}\indent A.B and A.A would like to acknowledge SINP, India and the Department of Atomic Energy (DAE), Government of India for research funding and Fellowship. M.R.H. acknowledges Science and Engineering Research Board (SERB), India for providing the financial support through National Post Doctoral Fellowship (NPDF) (Project No. PDF/2021/003366). I.D. thanks Technical Research Center, Department of Science and Technology (TRC-DST) for the support.

\bibliography{digestref}

\end{document}